\documentclass{revtex4}

\usepackage{graphicx}
\def\beq{\begin{equation}}
\def\eeq{\end{equation}}
\def\beqa{\begin{eqnarray}}
\def\eeqa{\end{eqnarray}}

\begin{document}
\bibliographystyle{revtex}

\title{High order corrections for top quark and jet production at the Tevatron}

\author{Nikolaos Kidonakis}
\altaffiliation{Currently at: Department of Physics, 
Southern Methodist University, Dallas, Texas 75275-0175}
\affiliation{Physics Department, Florida State University, 
Tallahassee, Florida 32306-4350}

\date{\today}

\begin{abstract}
An overview of the theoretical status of the top quark and 
single-jet inclusive production cross sections at the Tevatron is presented.
NLO results as well as NNLO-NNLL and higher order threshold expressions
derived from resummation calculations are discussed.
Numerical results are presented for top quark production at the Tevatron,
and it is shown that higher orders contribute
sizable corrections to the total cross section
and differential distributions, and they also dramatically reduce the 
factorization scale dependence of the cross section.
For jet production, the scale dependence is also reduced
but the NNLO threshold corrections are not very big.
\end{abstract}

\maketitle

\section{Introduction}

The calculation of cross sections in perturbative QCD relies on
the factorization of long-distance and short-distance physics.
The long-distance physics is described by parton distributions, $\phi$, 
which are determined from experiment, while the short-distance physics 
is embodied in hard-scattering factors, which are perturbatively calculable. 
The cross section for $t{\bar t}$ production in hadronic collisions
can then be written as
$\sigma_{h_1h_2\rightarrow t{\bar t}}
=\sum_f
\phi_{f_i/h_1} \otimes \phi_{f_j/h_2}
\otimes\hat{\sigma}_{f_i f_j\rightarrow t{\bar t}}$.
A similar equation holds for jet production.

Near threshold for the production of the $t{\bar t}$ pair (or jet) there is 
restricted phase space for real gluon emission. Thus, there is an
incomplete cancellation of infrared divergences 
between real and virtual graphs which results in large logarithms.
At $n$th order in $\alpha_s$, they are of the form
$[(\ln^m(s_4/m^2))/s_4]_+, \quad m \le 2n-1$,
with $s_4 \equiv s+t_1+u_1$. Note that $s_4 \rightarrow 0$ 
at threshold. Here $s,t_1,u_1$ are the
usual Mandelstam invariants.
These logarithms can be resummed to all orders in $\alpha_s$ by
factorizing soft gluons from the hard scattering \cite{NKGS,KOS98}.

\section{Top production}

At the Born level, the partonic channel $q{\bar q} \rightarrow t{\bar t}$
is the dominant production channel at the Tevatron;
it contributes over $90\%$ of the $t{\bar t}$ Born cross section.
The partonic channel $gg \rightarrow t{\bar t}$
contributes the remainder.
At NLO we have contributions from real gluon emission diagrams
and from one-loop virtual diagrams. We find sizable NLO corrections
for the $q {\bar q}$ channel and relatively large corrections
in the $gg$ channel.

The resummed heavy quark cross section in moment space, defined as
$\hat{\sigma}(N)=\int (ds_4/s) \, e^{-Ns_4/s} {\hat\sigma}(s_4)$,
with $N$ the moment variable,
is derived by refactorizing the cross section into hard
$H$ and soft $S$ functions that describe the hard scattering and 
soft gluon emission, respectively. 
We can then write the resummed cross section at NLL (next-to-leading 
logarithmic) or higher accuracy as
\beqa
{\hat{\sigma}}_{f_a f_b \rightarrow t{\bar t}}(N) =   
\exp\left[ E^{(f_a)}(N)+E^{(f_b)}(N)\right] \; 
{\rm Tr} \left \{H
\exp \left[\int_m^{m/N} \frac{d\mu}{\mu} \, \Gamma_S^{\dagger}\right] 
{\tilde S} \left(\alpha_s(m^2/N^2) \right)  
\exp \left[\int_m^{m/N} \frac{d\mu}{\mu} \, \Gamma_S \right] \right\}\, ,
\nonumber
\eeqa
where the incoming parton $N$-dependence is in $E^{(f_i)}(N)$,
the process-dependent functions $H$ and $S$ are matrices in the space of 
color exchanges where the trace is taken, and  
$\Gamma_S$ is the soft anomalous dimension 
matrix \cite{NKGS,KOS98}. Note that the plus distributions
$[(\ln^{2n-1}(s_4/m^2))/s_4]_+$ transform in moment space 
into $\ln^{2n}N$.

A prescription is needed to invert the moment-space resummed cross section
back to momentum space.
Alternatively we can expand the resummed cross section order by order in
$\alpha_s$ and match to NLO
to obtain NNLO-NNLL (next-to-next-to-leading order and 
next-to-next-to-leading logarithmic) and higher-order corrections 
\cite{NK00,KLMV}, 
thus avoiding prescription dependence and unphysical terms \cite{NK00}.
The problem with unphysical terms can be easily seen by studying the 
expansion at NLO. At NLO, the  soft plus virtual ($S+V$) corrections are
${\hat\sigma}_{q \bar q}^{(1)}(s_4)
=\sigma^B (\alpha_s/\pi) \{c_1 \delta(s_4)
 +c_2 [1/s_4]_+
+c_3 [(\ln(s_4/m^2))/s_4]_+\}$.
This result comes from inverting the moment space expression
${\hat\sigma}_{q \bar q}^{(1)}(N)
=\sigma^B (\alpha_s/\pi) \{c_1
 -c_2 \ln {\tilde N} +(c_3/2) (\ln^2 {\tilde N}+\zeta_2)\}$,
with ${\tilde N}=N e^{\gamma_E}$.
If we keep only NLL accuracy (as is the accuracy of the resummed cross 
section) in our results, i.e. only 
the $\ln^2 N$ and $\ln N$ terms, the inversion produces the 
following unphysical subleading terms: 
$\sigma^B (\alpha_s/\pi)[-(c_3/2) \gamma_E^2
+c_2 \gamma_E -(\zeta_2/2) c_3] \, \delta(s_4)$.
With $\sqrt{S}=1.8$ TeV and $m=\mu=175$ GeV/c$^2$,
the full NLO $S+V$ corrections are 1.26 pb.
If we keep only NLL accuracy they are 1.31 pb.
If we keep NLL accuracy but also include the unphysical subleading terms 
above, we find 0.39 pb, a result very far from the full $S+V$ corrections.
Thus at NLO, NNLO, and higher orders, we must not include unphysical 
subleading terms or use prescriptions that include such terms.

After expanding the resummed cross section in the $q{\bar q}$ channel
in the $\overline {\rm MS}$ scheme to two-loops, we find  
that the NLO threshold corrections agree with the exact NLO 
corrections in Ref. \cite{BNMSS}, while the new NNLO-NNLL 
corrections take the form 
\beqa
&&\hspace{-10mm}
s^2\frac{d^2{\hat \sigma}^{(2)}_{q{\bar q}\rightarrow 
t{\bar t}}(s_4,m^2,t_1,u_1)}{dt_1 \; du_1}
=\sigma^B_{q{\bar q}\rightarrow t{\bar t}} 
\left(\frac{\alpha_s(\mu_R^2)}{\pi}\right)^2 
\left\{8 C_F^2 \left[\frac{\ln^3(s_4/m^2)}{s_4}\right]_{+} 
+ C_F \left[\frac{\ln^2(s_4/m^2)}{s_4}\right]_{+} \right.
\nonumber \\ && \hspace{-15mm} \left.
{}\times \, \left[-\beta_0 
+12 \left({\rm Re}{\Gamma'}_{22}^{q\bar q}
-C_F+C_F\ln\left(\frac{sm^2}{t_1u_1}\right)
-C_F\ln\left(\frac{\mu_F^2}{m^2}\right)\right)\right]
+ C(s_4) \left[\frac{\ln(s_4/m^2)}{s_4}\right]_{+}\right\}
+{\cal O}\left(\left[\frac{1}{s_4}\right]_+\right)\,,
\nonumber
\eeqa
where $C(s_4)$ is a function that includes matching terms from the 
exact NLO calculation \cite{NK00,KLMV}.
Results have been derived through N$^4$LO in \cite{NK00}.
Analogous results have been derived for the $gg$ channel \cite{NK00,KLMV}.
Note that in addition we have derived all NNLO scale-dependent terms
for both channels \cite{KLMV}.

\begin{figure}
\resizebox{7cm}{5cm}{\includegraphics{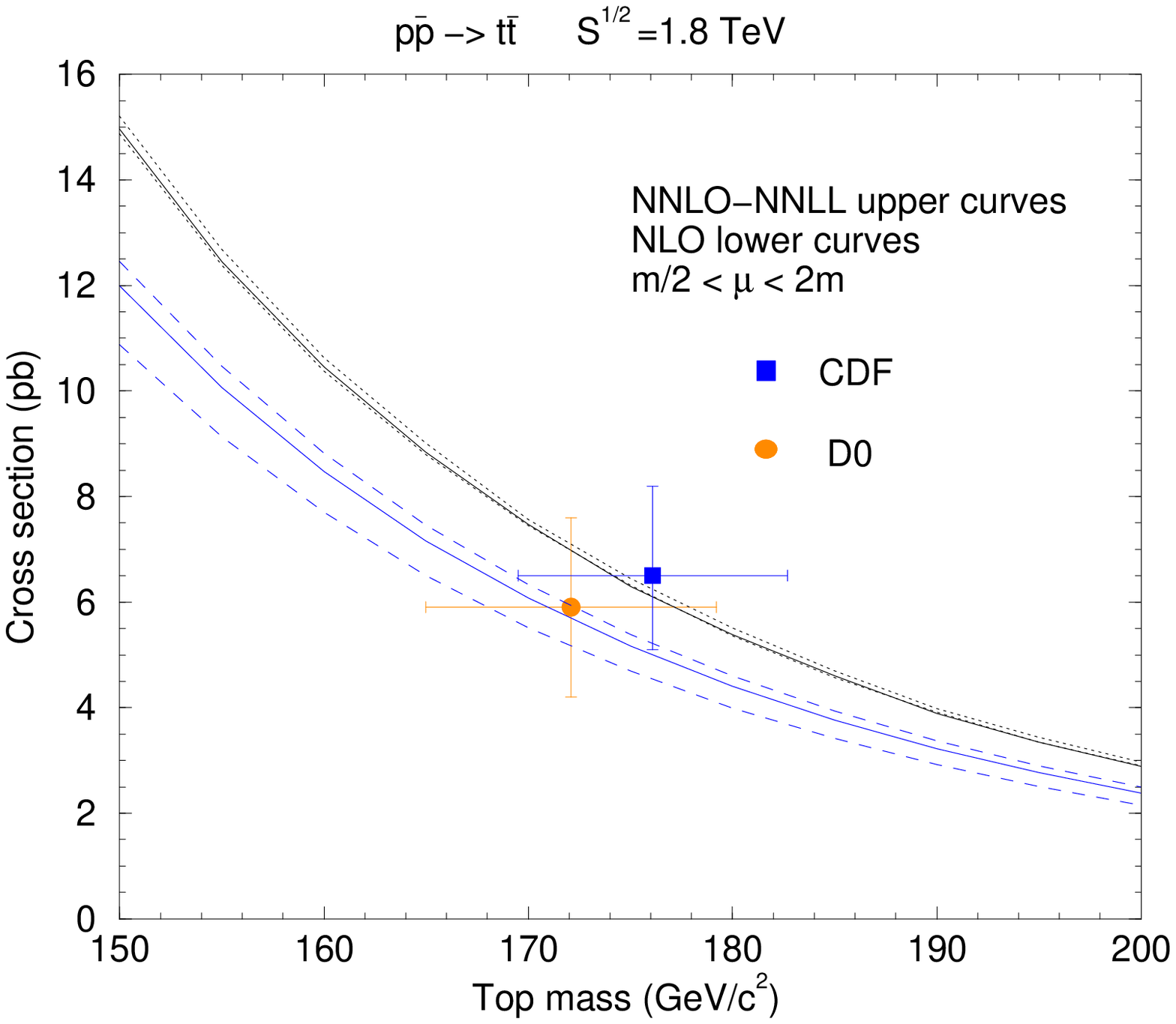}}
\hspace{5mm}
\resizebox{7cm}{5cm}{\includegraphics{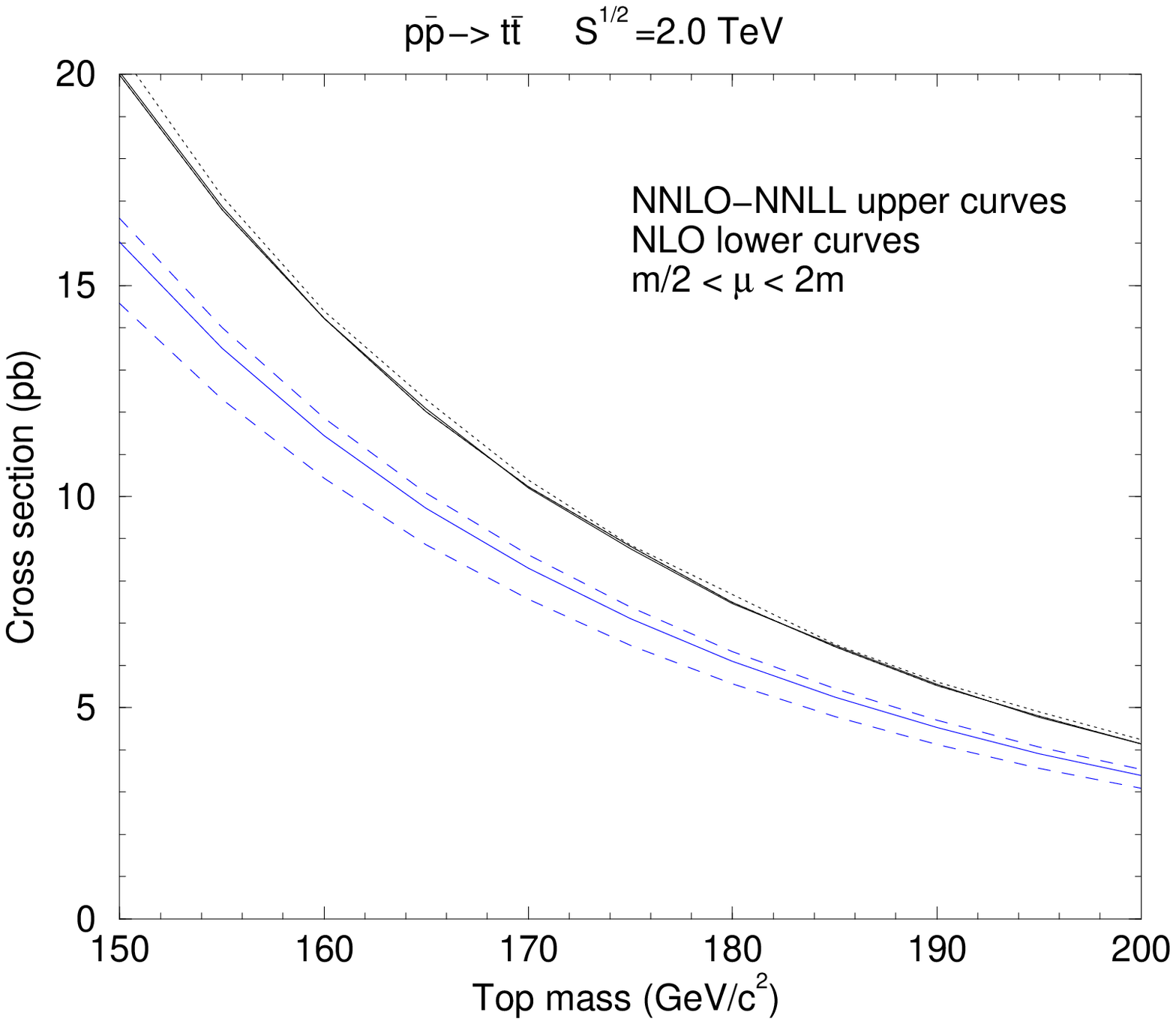}}
\caption{The total cross section for top quark production at the Tevatron
with $\sqrt{S}=1.8$ and $2.0$ TeV.}
\label{fig1}
\end{figure}

\begin{table}
\caption{Top cross section at the Tevatron}
\begin{tabular}{|c|c|c|} \hline
 $p{\bar p}\rightarrow t{\bar t}$  & $\sqrt{S}=1.8$ TeV 
& $\sqrt{S}=2.0$ TeV \\ \hline 
$\mu=\mu_F=\mu_R$   & NLO \quad NNLO & NLO  \quad NNLO \\ \hline
 $\mu=m/2$ & 5.4 \quad 6.4 & 7.4 \quad 8.9 \\ \hline
 $\mu=m$   & 5.2 \quad 6.3 & 7.1 \quad 8.8 \\ \hline
 $\mu=2m$  & 4.7 \quad 6.3 & 6.5 \quad 8.8 \\ \hline
\end{tabular}
\end{table}

In Table I and Fig. 1 we show values for the NLO and NNLO-NNLL $t{\bar t}$
cross section at the Tevatron for Run I and Run II. 
We find a sizable increase of the cross  
section, a dramatic decrease of the scale dependence $\mu$,
and good agreement with experiment, when we add the NNLO threshold corrections.

We have also investigated the uncertainties due to subleading logarithms 
(beyond NNLL accuracy) \cite{NK00}.
Including them as the largest error source in the calulation, we find
for $m=175$ GeV/$c^2$,
\beqa
  \sigma_{p{\bar p} \rightarrow t\bar{t}}(1.8 \; {\rm TeV})
= 6.3^{+0.1}_{-0.4} \;\;{\rm pb}\, ; \quad \quad
  \sigma_{p{\bar p} \rightarrow t\bar{t}}(2.0 \; {\rm TeV})
= 8.8^{+0.1}_{-0.5} \;\;{\rm pb}\, .
\nonumber
\eeqa

The above results are all for single-particle-inclusive (1PI) kinematics.
In pair-inclusive (PIM) kinematics the NNLO-NNLL corrections are 
smaller \cite{KLMV}.
If we take the average of the 1PI and PIM results, we find
\beqa
\sigma^{av}_{p{\bar p} \rightarrow t\bar{t}}(1.8 \; {\rm TeV})
= 5.8 \pm 0.4 \pm 0.1 \;\;{\rm pb} \, ; \quad \quad
  \sigma^{av}_{p{\bar p} \rightarrow t\bar{t}}(2.0 \; {\rm TeV})
= 8.0 \pm 0.6 \pm 0.1 \;\;{\rm pb}\, ,
\nonumber
\eeqa
where here the first error is due to the kinematics uncertainty
and the second due to the scale dependence.

Top transverse momentum distributions \cite{NKJStop} at NNLO-NNLL 
are given in \cite{NK00,DPF}. We also note that similar methods 
can be used for single-top production (including FCNC effects \cite{ABNK}).

\section{Jet production}

The invariant single-jet inclusive cross section involves contributions
from several parton-parton scattering subprocesses at lowest order:
$q_j{\bar q_j}  \rightarrow  q_j {\bar q_j}, \, 
q_j{\bar q_j}  \rightarrow  q_k {\bar q_k}, \, 
q_j{\bar q_k}  \rightarrow  q_j {\bar q_k}, \,  
q_j q_j  \rightarrow  q_j q_j, \,
q_j q_k  \rightarrow  q_j q_k, \, 
q {\bar q}  \rightarrow  g g, \,
g g  \rightarrow  q {\bar q}, \, 
q g  \rightarrow  q g, \,$ and  
$g g  \rightarrow   g g$.

We resum QCD corrections in high-$E_T$ jet production 
to NLL accuracy \cite{KOS98}
and expand the resummed cross section to NNLO \cite{NKJO2}.
There are additional complications relative
to $t {\bar t}$ production because of the final-state jets.

We have calculated the NNLO-NLL corrections for all partonic subprocesses.
For example the NNLO threshold corrections for 
$q_j {\bar q}_k \rightarrow q_j {\bar q}_k $ are
\beqa
E_J\frac{d^3{\hat \sigma}^{(2)}_{q_j {\bar q}_k \rightarrow q_j {\bar q}_k}}
{d^3p_J}&=&\left(\frac{\alpha_s}{\pi}\right)^2
\sigma^B_{q_j {\bar q}_k \rightarrow q_j {\bar q}_k}
\left\{2C_F^2 \left[\frac{\ln^3(s_4/p_T^2)}{s_4}\right]_+ 
+3C_F \left[\frac{\ln^2(s_4/p_T^2)}{s_4}\right]_+ \right.
\nonumber \\ &&  \hspace{-45mm} \left.
{}\times \, \left[-2C_F\ln\left(\frac{\mu_F^2}{p_T^2}\right)
-\frac{3}{2}C_F-\frac{(N_c^2+1)}{N_c}\ln\left(\frac{-t}{s}\right)
-\frac{(N_c^2-5)}{N_c}\ln\left(\frac{-u}{s}\right)-\frac{\beta_0}{12}\right]
\right\}
+{\cal O}\left(\left[\frac{\ln(s_4/p_T^2)}{s_4}\right]_+\right), 
\nonumber
\eeqa
with $p_T$ the jet's transverse momentum.
Numerical results for jet production at the Tevatron are given
in Fig. 2. The NNLO threshold corrections decrease the scale dependence,
but they do not provide a sizable increase of the cross section, especially
for smaller values of the scale.

\begin{figure}
\resizebox{7cm}{5cm}{\includegraphics{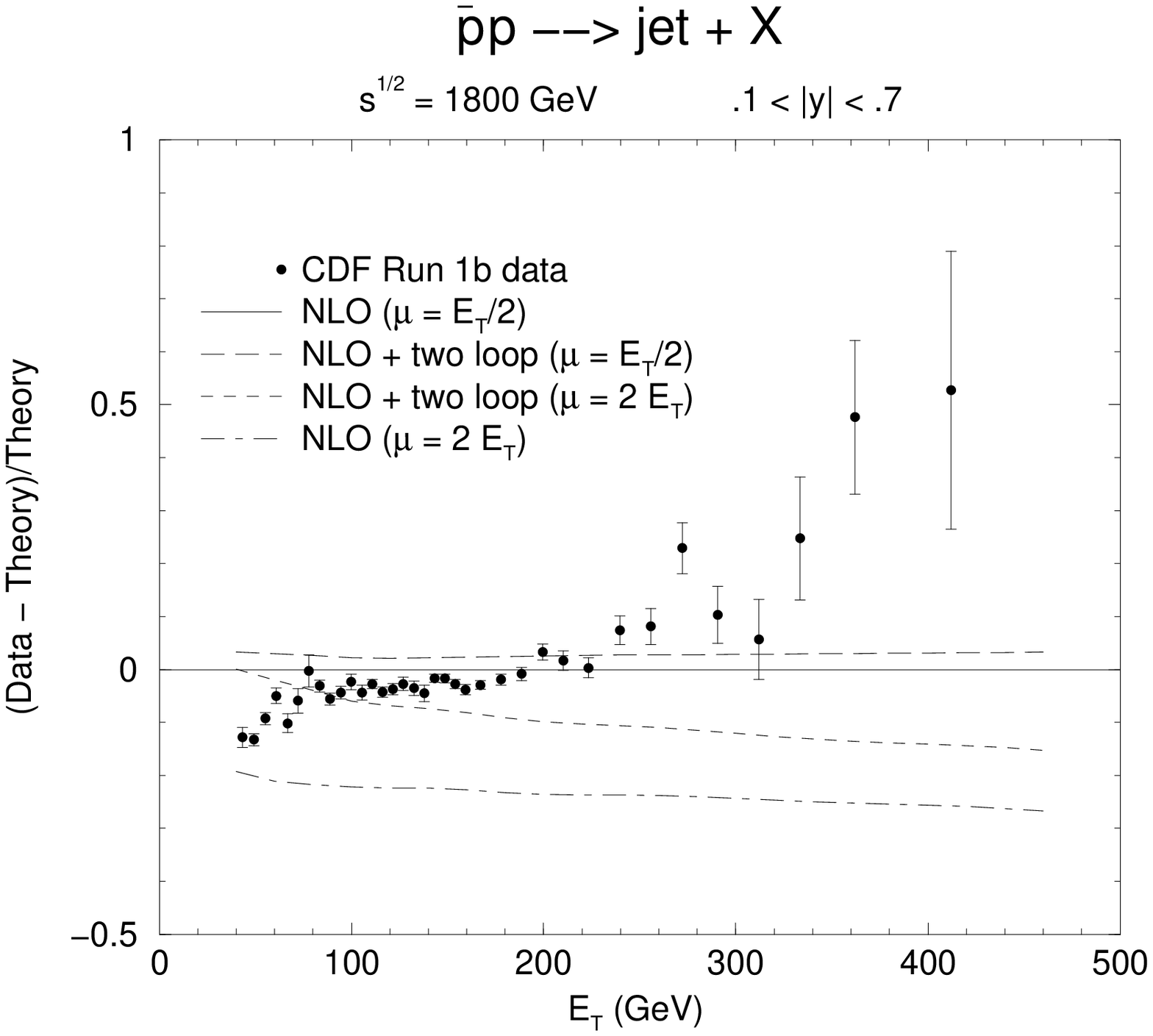}}
\hspace{5mm}
\resizebox{7cm}{5cm}{\includegraphics{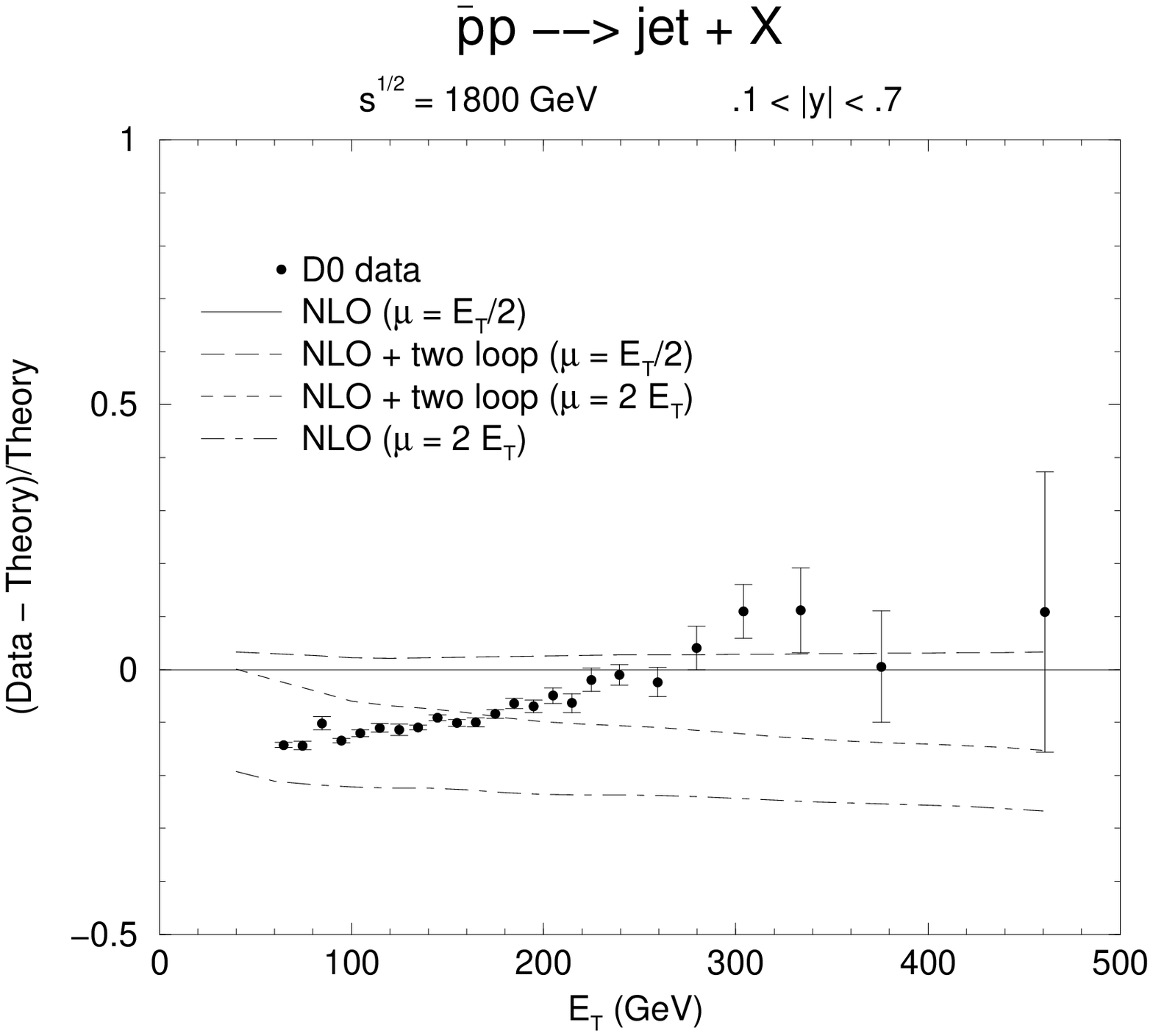}}
\caption{Jet production cross section at the Tevatron at Run I 
(${\sqrt S}=1.8$ TeV) versus CDF and D0 data.}
\label{fig2}
\end{figure}

\section{Conclusions}

Threshold resummation for heavy quark and jet production
is a very powerful formalism and it allows us to derive
high-order corrections to the cross sections for these processes.
The resummed cross section has been expanded analytically through
N$^4$LO and care has been taken to exclude unphysical subleading terms.

The NNLO-NNLL (and higher-order) corrections 
provide new analytical and numerical predictions
for the total cross section and differential distributions 
for both top quark and jet production at the Tevatron.
We note a significantly reduced scale dependence at higher orders
for both processes,
and a significant increase of the $t {\bar t}$ cross section after
resummation.

\end{document}